\documentclass[]{emulateapj-rtx4}

\usepackage{natbib,color} 
\bibpunct{(}{)}{;}{a}{}{,}

\DeclareRobustCommand{\ion}[2]{%
\relax\ifmmode
\ifx\testbx\f@series
{\mathbf{#1\,\mathsc{#2}}}\else
{\mathrm{#1\,\mathsc{#2}}}\fi
\else\textup{#1\,{\mdseries\textsc{#2}}}%
\fi}

\shorttitle{Magnetic Properties and Flow Angle of the Inverse Evershed Flow}
\shortauthors{Beck, C.; Choudhary, D.P.}

\begin{document}
\title{Magnetic Properties and Flow Angle of the Inverse Evershed Flow at Its Downflow Points}


\author{C. Beck}
\affil{National Solar Observatory (NSO)}
\author{D.P. Choudhary}
\affil{ California State University (CSUN)}


%\altaffiltext{1}{Departamento de Astrof\'isica, Universidad de La Laguna, E-38205 La Laguna, Tenerife, Spain}


\begin{abstract}
We determined the direction and strength of the photospheric and lower chromospheric magnetic field in the umbra and penumbra of a sunspot from inversions of spectropolarimetric observations of photospheric lines at 617\,nm and 1565\,nm, and the chromospheric \ion{Ca}{ii} IR line at 854\,nm, respectively. We compare the magnetic field vector with the direction of 75 flow channels that harbor the chromospheric inverse Evershed effect (IEF) near their downflow points (DFPs) in the sunspot's penumbra. The azimuth and inclination of the IEF channels to the line of sight (LOS) were derived from spatial maps of the LOS velocity and line-core intensity of the \ion{Ca}{ii} IR line and a thermal inversion of the \ion{Ca}{ii} IR spectra to obtain temperature cubes. We find that the flow direction of the IEF near the DFPs is aligned with the photospheric magnetic field to within about $\pm$\,15\,deg. The IEF flow fibrils make an angle of 30--90\,deg to the local vertical with an average value of about 65\,deg. The average field strength at the DFPs is about 1.3\,kG. Our findings suggest that the IEF in the lower chromosphere is a field-aligned siphon flow, where the larger field strength at the inner footpoints together with the lower temperature in the penumbra causes the necessary gas pressure difference relative to the outer footpoints in the hotter quiet Sun with lower magnetic field strength. The IEF connects to magnetic field lines that are not horizontal like for the regular photospheric Evershed flow, but which continue upwards into the chromosphere indicating an "uncombed" penumbral structure. 
\end{abstract}

\section{Introduction}
The magnetic topology in sunspots on the solar surface shows an increase in its complexity from the central dark umbra towards the outer end of the penumbra. While umbral magnetic fields are barely inclined relative to the local surface normal and primarily expand into the upper solar atmosphere, the magnetic field lines in the penumbra show a mixture of different inclination angles \citep[\textit{e.g.},][]{solanki2003,bellot+etal2004,beck2008}. Some of the magnetic field lines return back to the photosphere within the penumbra \citep{deltoroiniesta+etal2001,franz+schliche2013,ruizcobo+asensio2013,pozuelo+etal2016}, while some others continue upwards and outwards into the so-called superpenumbra that is seen in chromospheric diagnostics to extend beyond the sunspot boundary visible in the photosphere \citep{joshi+etal2017}. 

The more horizontal penumbral filaments carry the regular photospheric Evershed effect \citep[\textit{e.g.},][]{evershed1909,reza+etal2006,khomenko+etal2015}, a radial outflow from the outer umbral towards the outer penumbral boundary that even continues into the moat region surrounding the sunspot \citep{reza+etal2006}. The photospheric mass motions in and outside of sunspots are a basic aspect of their evolution related to, for instance, the energy transport in the penumbra or the loss of magnetic flux through moving magnetic features \citep{harvey+harvey1973,vrabec1974,cabrerasolana+etal2006,kubo+etal2008,rempel2015}. 

The chromospheric inverse Evershed flow (IEF) goes in the reverse direction towards the umbra along fibrils that join the sunspot and the outer superpenumbral boundary. Its connection to the topology of the magnetic field in the penumbra is less well known. While different mechanisms have been proposed as the driver of the regular Evershed effect \citep{montesinos+thomas1997,rempel2012,siutapia+etal2017}, the IEF is commonly assumed to be caused by a siphon flow \citep[\textit{e.g.},][]{thomas1988} between an inner footpoint in the strong magnetic field inside a sunspot and an outer footpoint in a region with lower magnetic field strength. The different value of the magnetic pressure term at the two footpoints and the low temperature in the penumbra lead to a gas pressure difference directed towards the inner footpoint that can drive a flow along the connecting magnetic field lines. 

\begin{figure*}
\centerline{\resizebox{6.5cm}{!}{\includegraphics{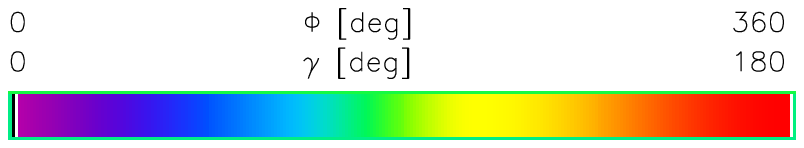}}\hspace*{.5cm}\resizebox{6.5cm}{!}{\includegraphics{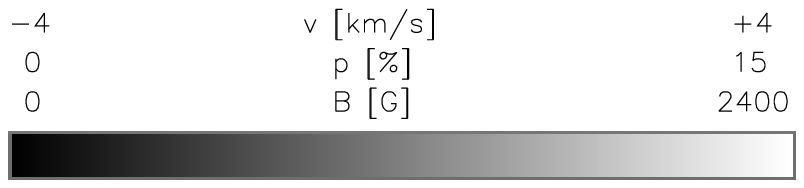}}}$ $\\
\centerline{\resizebox{17cm}{!}{\hspace*{1cm}\includegraphics{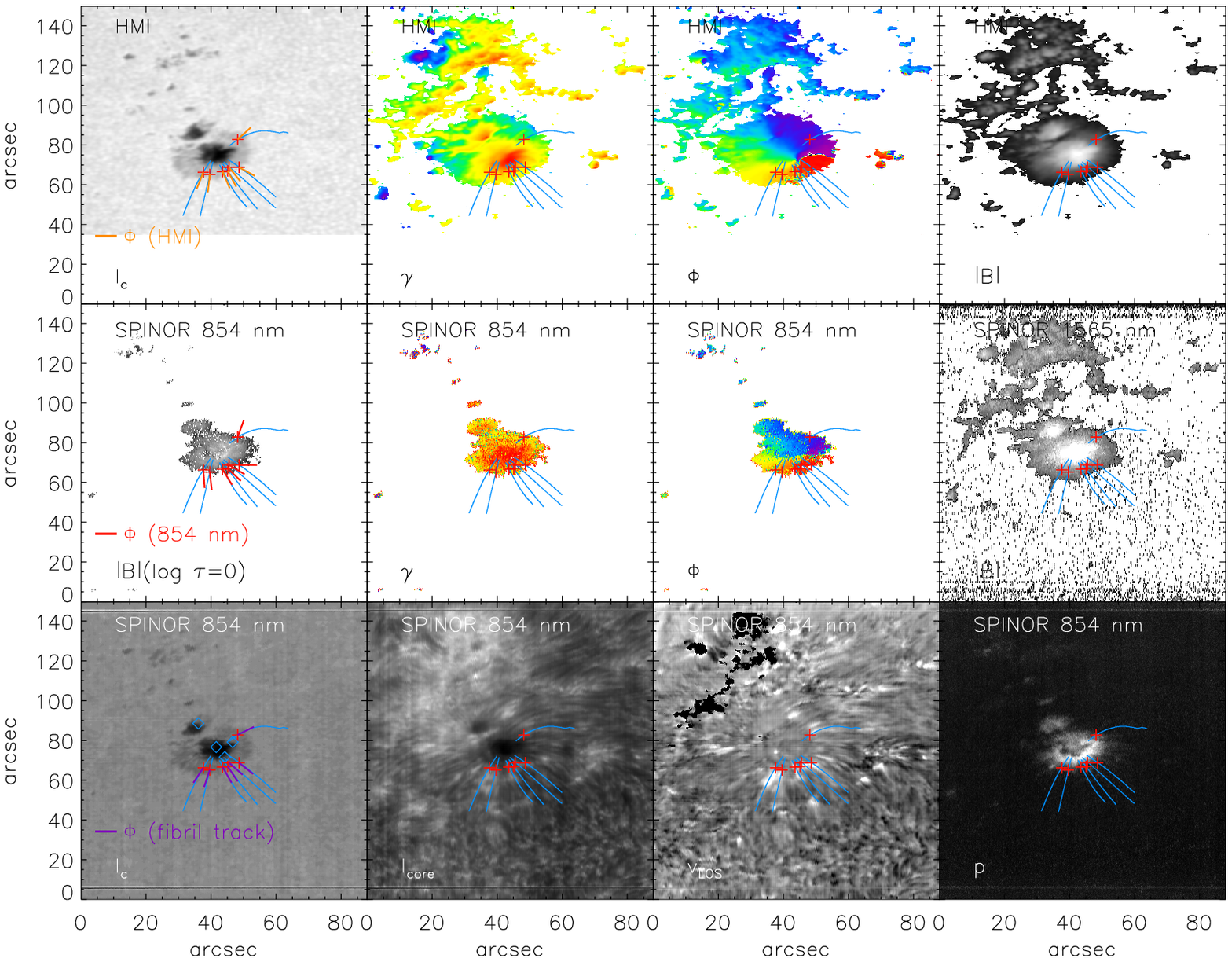}}}$ $\\$ $\\$ $\\
\caption{Overview maps of the first scan of NOAA 11809 at UT 15:24 on 03 Aug 2013. Bottom row, left to right: continuum intensity $I_c$, line-core intensity $I_{core}$, line-core velocity $v_{LOS}$ and polarization degree $p$ from the SPINOR 854\,nm spectra. Middle row, left to right: magnetic field strength $B$ at log\,$\tau = 0$, inclination $\gamma$ and azimuth $\phi$ from the inversion of the SPINOR 854\,nm spectra, and $B$ from the inversion of the SPINOR 1565\,nm spectra. Top row, left to right: $I_c$, $\gamma$, $\phi$ and $B$ from the inversion of the HMI data. All values of $\gamma$ and $\phi$ are in the line-of-sight reference frame. The blue curved lines indicate the tracks of flow fibrils. The red pluses indicate the inner footpoints where the field lines return to the photosphere. The purple, red and orange bars in the first column indicate the direction of the fibril tracks and the azimuth of the 854\,nm and HMI inversion at the footpoints, respectively. The four blue diamonds in the lower left panel indicate the locations of the \ion{Ca}{ii} IR spectra shown in Figure \ref{spec_exam}.}\label{fig1}
\end{figure*}

The magnetic field orientation in the solar atmosphere can be derived with spatial resolution from an analysis of the polarization signal of individual profiles of Zeeman-sensitive spectral lines \citep{deltoroiniesta+ruizcobo2016}. A direct determination of the direction of mass flows is not possible in the same way without additional assumptions on the symmetries of the flow pattern \citep{schliche+schmidt2000,bellot+etal2003}. While the apparent direction of flow channels in the plane perpendicular to the line of sight (LOS) can be determined from spatial maps of, \textit{e.g.}, the LOS velocity of spectral lines \citep{choudhary+beck2018}, the flow angle to the LOS cannot be directly derived from individual spectra because only the projection of the true flow velocity to the LOS is known. This obstacle can be circumvented to some extent by a thermal inversion, \textit{i.e.}, the determination of the temperature structure along the LOS. If the thermal structure traces the flow channels, also their direction relative to the LOS can be determined from an analysis of such temperature cubes or two-dimensional temperature slices \citep{beck+etal2014}. In this study, we investigate the flow angle of the IEF derived by this approach and compare its properties to those of the magnetic field vector at the downflow points of the IEF in the penumbra.

Section \ref{secobs} describes the observational data used whose analysis is explained in Section \ref{analysis}. Section \ref{secres} gives the results on magnetic field properties and flow angle obtained, which are discussed in Section \ref{secdisc}. Section \ref{secconcl} provides our conclusions.
%summarized in Section \ref{secsumm} and
\section{Observations}\label{secobs}
The ground-based observations used here are described in detail in \citet[][in the following Paper I]{choudhary+beck2018}, so we only repeat their main characteristics. We observed the leading sunspot of the active region NOAA 11809 on 03 August 2013 with the \textit{SPectropolarimeter for Infrared and Optical Regions} \citep[SPINOR;][]{socasnavarro+etal2006} at the Dunn Solar Telescope (DST). We obtained in total nine maps of the sunspot from UT 15:24 to UT 18:43 in the chromospheric \ion{Ca}{ii} infrared (IR) line at 854.2\,nm and in a wavelength region around 1565\,nm that contains several photospheric \ion{Fe}{i} lines (see Table \ref{tablines}). The spatial (spectral) sampling along the slit was 0\farcs36 (5.5\,pm) at 854\,nm and 0\farcs55 (20.6\,pm) at 1565\,nm. The field of view (FOV) along the slit was about 150$^{\prime\prime}$\,\footnote{The 90$^{\prime\prime}$ for \ion{Ca}{ii} IR given in Paper I was wrong.}, while the spatial extent scanned was 400 (200) steps of 0\farcs22 step width for the first (all other) maps.

These observations are complemented here by co-aligned data from the Helioseismic and Magnetic Imager \citep[HMI;][]{scherrer+etal2012} on-board the Solar Dynamics Observatory \citep[SDO;][]{pesnell+etal2012}. We use the results of an inversion of HMI data taken on 03 Aug 2013 with the Very Fast Inversion of the Stokes Vector code \citep{borrero+etal2011} courtesy of R. Rezaei done with the settings described in \citet{kiess+etal2014}. Because of a miscommunication on the observing time (local {\emph{vs}}.~universal time), the closest HMI inversion available was taken at UT 12:36, {\emph{i.e.}}, three hours before the observations at the DST, but the sunspot did not significantly evolve during that time span (see Figure \ref{fig1}). 

\begin{table}
\caption{Photospheric \ion{Fe}{i} lines around 1565\,nm. EP = excitation potential. $\alpha, \sigma$ = broadening parameters.}\label{tablines}
\begin{tabular}{cccccc}
 $\lambda$ & EP & log gf & transition & $\alpha$ & $\sigma$ \cr
  nm & eV &\cr\hline\hline
 1558.8264 &   6.366  & 0.2  &  5D 4.0- 5D 4.0 & 0  &   0 \cr
 1562.1658 &  5.539  &0.3   & 5D 4.0- 5D 4.0  &0   &  0   \cr
 1563.1950 &  5.352  &0.15  &  7D 4.0- 7D 4.0 & 0  &   0   \cr
 1564.5020 &   6.311  &-0.45 &  7P 2.0- 7P 2.0 & 0.291 &3.36e-14\cr 
 1564.8515 &    5.426 & -0.669&  7D 1.0- 7D 1.0&  0.229 & 2.74e-14\cr
 1565.2874 &   6.246  &-0.095 & 7D 5.0- 7D 4.0 & 0.330  &4.00e-14\cr
 1566.2018 &    5.829 &  0.19 &  5F 5.0- 5F 4.0&  0.240 &3.36e-14\cr
 1566.5245 &   5.978  &-0.42  & 5F 1.0- 5D 1.0 & 0.230 &3.59e-14\cr
\end{tabular}
\end{table}

\section{Data analysis}\label{analysis}

\subsection{Inversion of Spectropolarimetric \ion{Fe}{i} Data at 1565\,nm}
The spectropolarimetric observations at 1565\,nm were analyzed with the Stokes Inversion based on Response functions code \citep[SIR;][]{cobo+toroiniesta1992}. We included the eight \ion{Fe}{i} lines listed in Table \ref{tablines} that were covered in each spectrum. The inversion setup used a variable stray-light contribution, a single magnetic component in the umbra, two magnetic components in the penumbra, and a magnetic and a field-free component for profiles with significant polarization signal in the quiet Sun. All magnetic field parameters (inclination $\gamma$, azimuth $\phi$, field strength $B$) were assumed to be constant with optical depth, while the temperature stratification was modified using two nodes. The 1565\,nm inversion results turned out to be somewhat noisy especially in $\gamma$ and $\phi$ because the spectral sampling of 20\,pm partially under-sampled the lines. We thus decided to only use the field strength value $B$ from this inversion (rightmost panel in the middle row of Figure \ref{fig1}) that was calculated as the average value of the two magnetic inversion components weighted with their relative fill factor in the penumbra. 

\begin{figure}
$ $\\
\centerline{\resizebox{7.5cm}{!}{\includegraphics{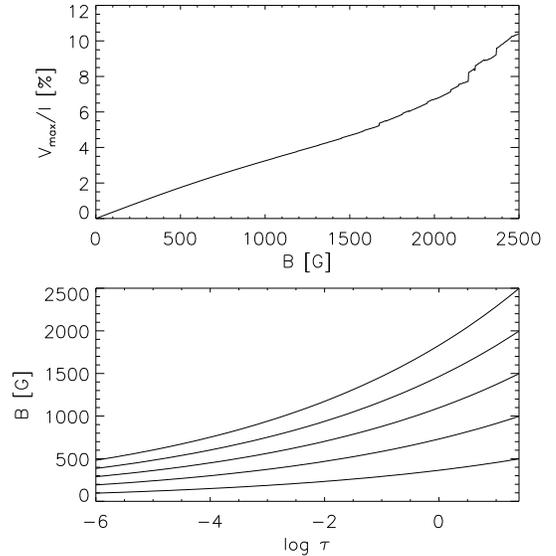}}}$ $\\
\caption{Derivation of initial values for the magnetic field strength in the fit of the SPINOR 854\,nm spectra. Bottom: magnetic field stratifications for a field strength $B_0$ of 0.5--2.5\,kG at log $\tau$ = +1.4 with an exponential decay constant of $\Delta\tau =4.5$. Top: resulting maximal Stokes $V$ amplitude of \ion{Ca}{ii} IR at 854\,nm as a function of $B_0$.}\label{fig_calcurves}
\end{figure}
\begin{figure}
$ $\\$ $\\
\begin{minipage}{4.65cm}
\resizebox{4.65cm}{!}{\includegraphics{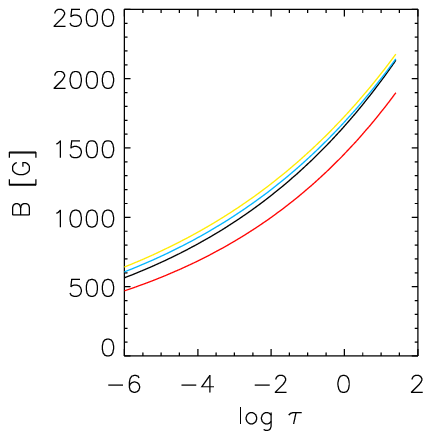}}
\end{minipage}\hspace*{.5cm}
\begin{minipage}{3.25cm}
\resizebox{3.25cm}{!}{\includegraphics{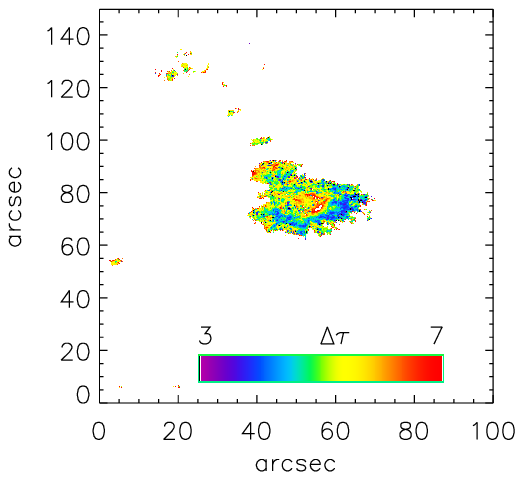}}\\$ $\\
\end{minipage}
\caption{Magnetic field stratifications for the spectra shown in Figure \ref{spec_exam} (left panel) and exponential decay constant $\Delta\tau$ across the FOV of the first scan (right panel). }\label{mag_exam}
\end{figure}
\subsection{Inversion of Spectropolarimetric \ion{Ca}{ii} IR Data at 854\,nm}
The intensity spectra of \ion{Ca}{ii} IR at 854\,nm were inverted with the CAlcium Inversion using a Spectral ARchive code \citep[CAISAR;][see also Paper I]{beck+etal2013,beck+etal2015}. For the current study, we extended this code to a full-Stokes inversion code based again on SIR to derive the magnetic field vector from the \ion{Ca}{ii} IR spectra. 

\begin{figure*}
$ $\\
\resizebox{17.6cm}{!}{\includegraphics{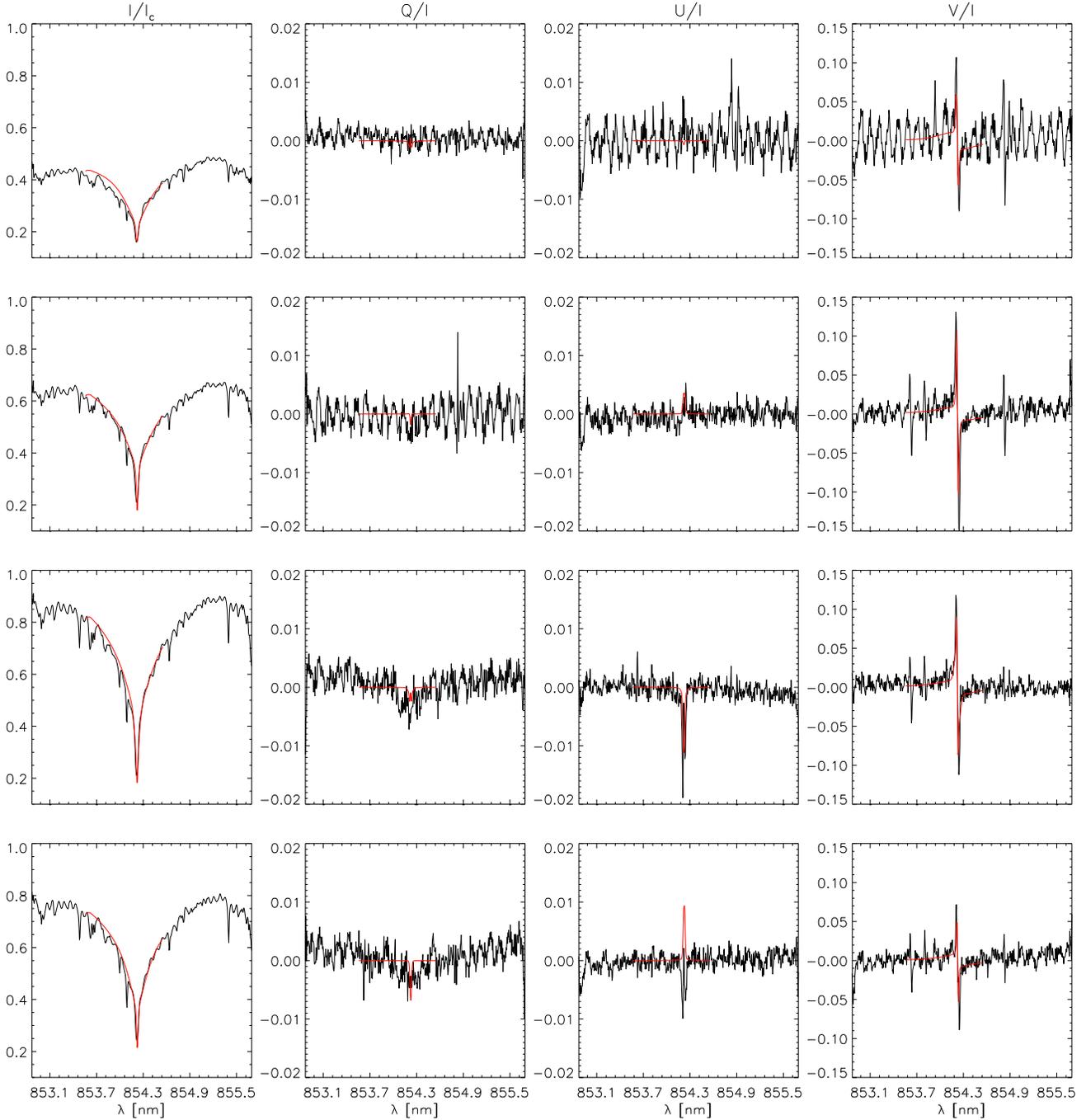}}\\$ $\\
\caption{Example fits of (left to right) Stokes $IQUV$ spectra of the \ion{Ca}{ii} IR at 854\,nm inversion. Black lines indicate the observations and red lines the best-fit result. The four spectra were located from top to bottom in the umbra, the inner and mid penumbra, and the largest pore in the FOV.}\label{spec_exam}
\end{figure*}

\begin{figure*}
\centerline{\resizebox{5.5cm}{!}{\includegraphics{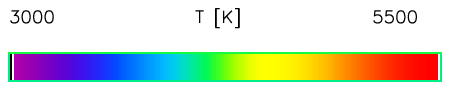}}}$ $\\
\centerline{\resizebox{11.cm}{!}{\includegraphics{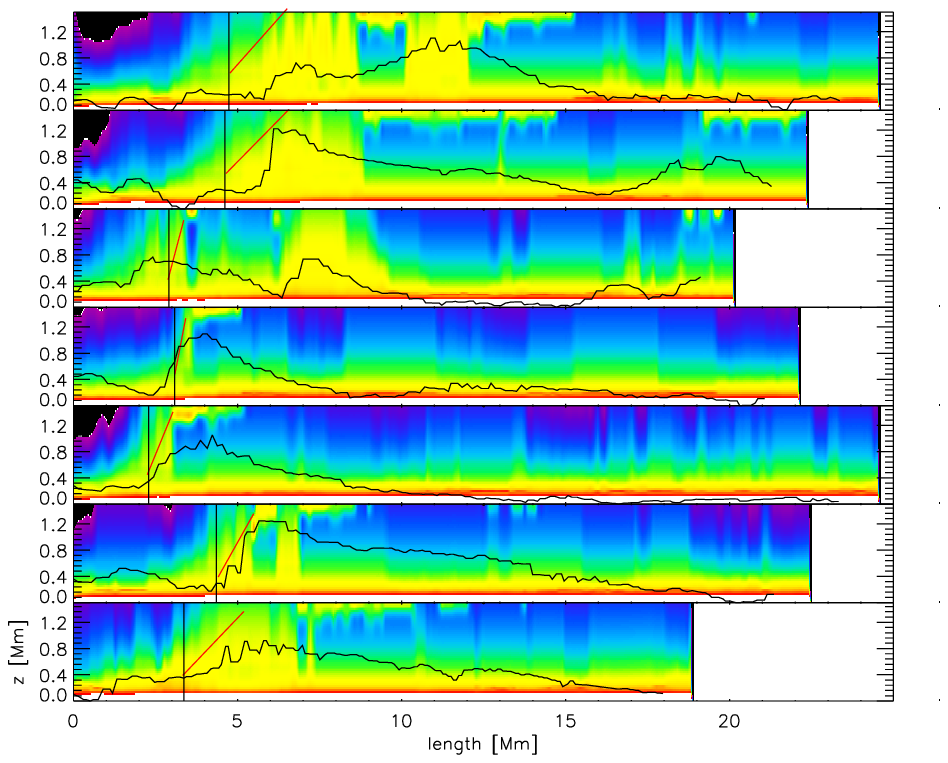}}}$ $\\$ $\\$ $\\
\caption{Determination of the flow angle from the CAISAR temperature results along the fibril tracks shown in Figure \ref{fig1}. The images in the background show the temperature along the fibrils. The inclined red lines indicate the manually determined flow angle. The vertical black lines indicate the inner lower footpoint. The LOS velocity along the fibrils is over-plotted as black lines in arbitrary units for comparison.}\label{fig2}
\end{figure*}

We first created two calibration curves to obtain suited initial values for the inversion to speed up the convergence. We synthesized Stokes $IQUV$ spectra of the \ion{Ca}{ii} IR line at 854\,nm with a field strength of 300\,G and a varying magnetic field inclination using the Harvard-Smithsonian Reference Atmosphere \citep[HSRA;][]{gingerich+etal1971} as the temperature stratification. We calculated the ratio of linear to circular polarization
\begin{eqnarray}
L/V = \frac{\sqrt{Q^2 + U^2}}{V} \propto 0.5 \frac{\sin^2 \gamma}{\cos \gamma} \,.
\end{eqnarray}
around the wavelength of maximal Stokes $V$ amplitude in the synthetic spectra which is directly related to the inclination $\gamma$ in the weak-field approximation \citep[see, \textit{e.g.},][]{jefferies+etal1989}. For observations, the ratio $L/V$ can be easily determined in the same way and then either be directly converted to $\gamma$ through \citep{beckthesis2006}
\begin{eqnarray}
\cos \gamma = -L/V + \sqrt{(L/V)^2 +1}
\end{eqnarray}
or compared with the $L/V$ calibration curve to retrieve $\gamma$. It turned out that the direct conversion is as accurate for the \ion{Ca}{ii} IR line at 854\,nm because of its small Land{\'e} coefficient.

% \citep[see also][]{centeno2018}.

To obtain an initial value for the field strength, we generated synthetic spectra using the HSRA for temperature, an inclination of 0\,deg, and magnetic field stratifications of the shape
\begin{eqnarray}
B(\log \tau) = B_0 \exp^{-\frac{\log \tau}{\Delta\tau}}
\end{eqnarray}
for field strengths $B_0$ of 0--2.5\,kG with $\Delta\tau = 4.5$. We calculated the maximal Stokes $V$ amplitude of the synthetic spectra to obtain a calibration curve of $V$ as a function of $B_0$ (see the top panel of Figure \ref{fig_calcurves}). The initial value of the magnetic field azimuth $\phi$ was derived from $\tan 2 \phi = Q/U$.

\begin{figure*}
\resizebox{8.8cm}{!}{\includegraphics{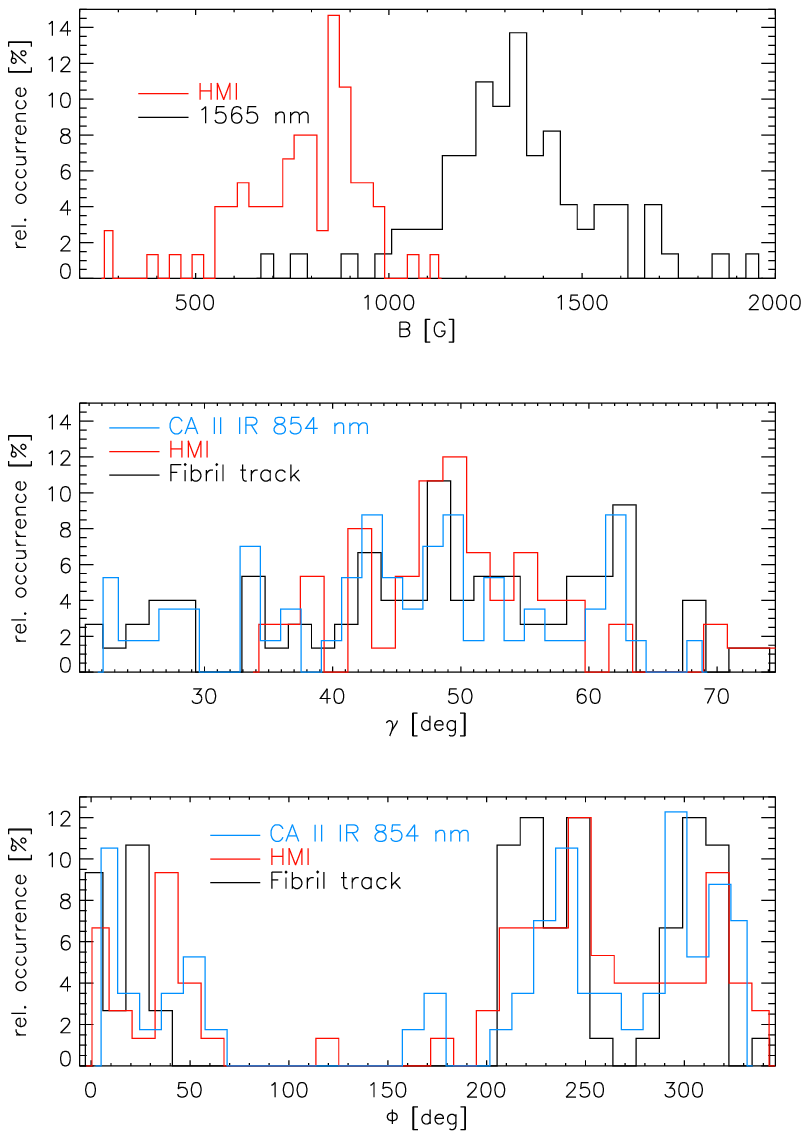}}\resizebox{8.8cm}{!}{\includegraphics{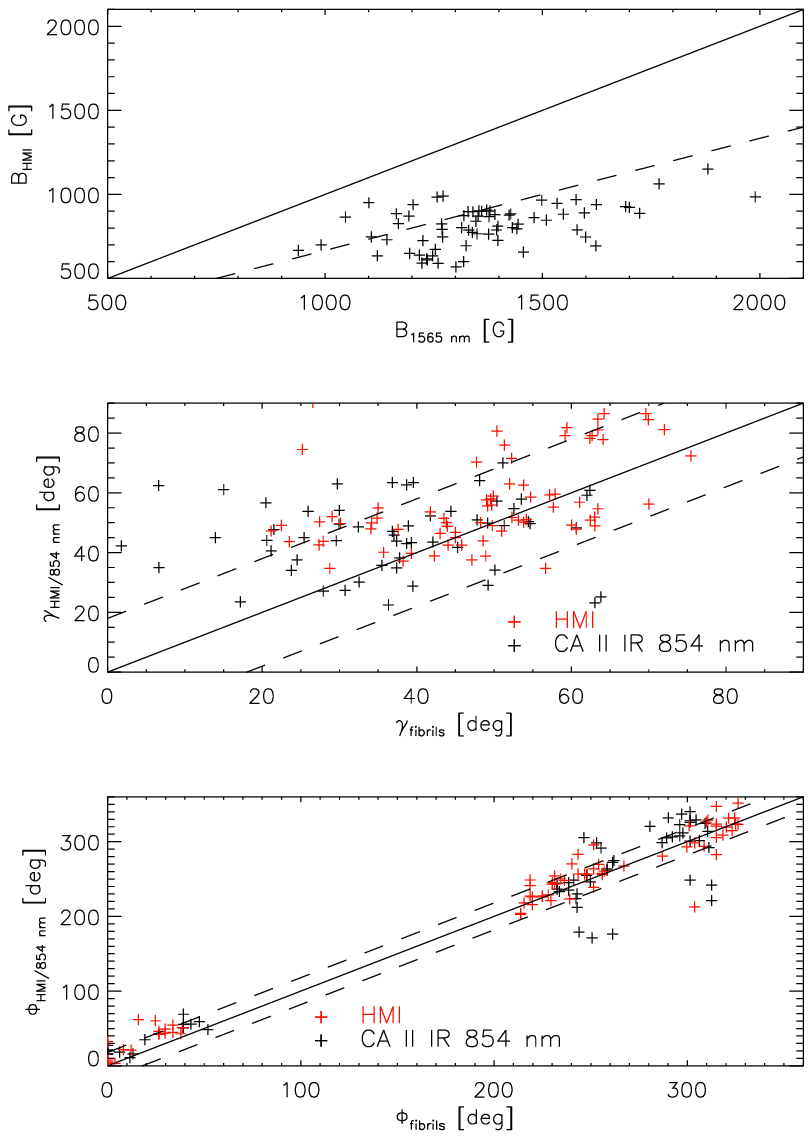}}
\caption{Left column: histograms of properties at the photospheric footpoints. Top panel: magnetic field strength $B$ from the 1565\,nm inversion (black line) and HMI (red line). Middle panel: inclination $\gamma$ from the temperature cuts along the fibrils (black line), HMI magnetic field inclination (red line) and the same from the \ion{Ca}{ii} IR inversion (blue line). Bottom panel: azimuth angle $\phi$ from the fibril tracks (black line), HMI magnetic field azimuth (red line) and the same from the \ion{Ca}{ii} IR inversion (blue line). Right column: scatter plots of properties at the photospheric footpoints. Top panel: magnetic field strength $B$ from HMI {\emph{vs}}.~the 1565\,nm inversion. The solid black line indicates a one-to-one correlation. The dashed line corresponds to a slope of 2/3. Middle panel: magnetic field inclination from the \ion{Ca}{ii} IR inversion (black pluses) and HMI (red pluses) {\emph{vs}}.~the inclination from the temperature cuts along the fibrils. Bottom panel: field azimuth from the \ion{Ca}{ii} IR inversion (black pluses) and HMI (red pluses) {\emph{vs}}.~the azimuth derived from the fibril tracks. The dashed lines denote a range of $\pm 18$\,deg around the one-to-one correlation. }\label{histo1}
\end{figure*}

For the inversion of the spectra, we then ran a standard iterative least-squares minimization by a gradient method with the free parameters $\gamma, \phi, \Delta\tau$ and $B_0$ using SIR to create the synthetic spectra. The temperature stratification for each profile was taken from the CAISAR inversion results of the Stokes $I$ spectra and kept fixed, since it already provides a best-fit to the observed intensity spectrum. We used a non-standard definition of $\chi^2$ given by
\begin{eqnarray}
\chi^2 &=& \frac{1}{w_Q}  ( Q_{max}^{obs} - Q_{max}^{synth})^2 + \frac{1}{w_U}  ( U_{max}^{obs} - U_{max}^{synth})^2 + \cr\nonumber 
&+& \frac{1}{w_{V1}}  ( V_{max}^{obs} - V_{max}^{synth})^2 + \frac{1}{w_{V2}} \int ( V^{obs} - V^{synth} )^2 d\lambda \,,
\end{eqnarray}
where $QUV_{max}$ are the values of the largest $QUV$ amplitudes in the profiles maintaining their sign and the $\frac{1}{w_i}$ are weighting coefficients. 

This choice of $\chi^2$ weights fitting the values of $\gamma$ and $\phi$ more than $B$ or $\Delta\tau$ and was selected because in the current study we are primarily interested in the direction of the magnetic field vector. The fit was limited to 40 iterations of the gradient determination with a mean duration of 4\,s {\emph{per}} profile and 8\,s at maximum. 
%The speed of 0.1\,s {\emph{per}} intensity profile of the CAISAR inversion is negligible in comparison.

The first three panels in the middle row of Figure \ref{fig1} show $B, \gamma$ and $\phi$ from the 854\,nm inversion of the first scan. Figure \ref{spec_exam} shows four randomly picked examples of $IQUV$ profiles in the first scan that all show a satisfactory fit, while the left panel of Figure \ref{mag_exam} displays the corresponding magnetic field stratifications. The right panel of Figure \ref{mag_exam} shows the variation of the exponential decay constant $\Delta\tau$ across the FOV in the \ion{Ca}{ii} IR inversion. It shows a systematic spatial variation with larger values in the umbra than in the penumbra. The coherent spatial evolution of $\Delta\tau$ implies that there is no simple trade-off between $B$ and $\Delta\tau$ that would lead to a salt-and-pepper pattern. 
%Given the LTE limitation in the inversion and the fact that we currently focus on the field direction, we will postpone any more detailed investigation of the full-Stokes \ion{Ca}{ii} IR inversion to a later study.      
\subsection{Fibril Tracks, Thermodynamic Parameters and Flow Angle}
From the analysis done in Paper I, we already obtained the spatial tracks of the flow fibrils and several thermodynamic parameters such as the line-core velocity, line-core intensity and the temperature stratifications along the fibril tracks. For the current study, we additionally determined the flow direction. 

We used the temperature stratifications along the fibril tracks to manually define the flow angle relative to the LOS (see Figure \ref{fig2}). We re-sampled the temperature stratifications first on an equidistant grid in km along the fibril tracks and then converted the optical depth scale in units of $\log \tau$ on the vertical axis to geometrical height $z$ assuming the same relation between $z$ and $\log \tau$ as in the HSRA model. We marked two points, one at photospheric level and one at chromospheric heights, for each fibril track to trace the slope of the temperature enhancements near the downflow points. We only used a subset of the fibril tracks in each scan of the sunspot where this definition was feasible, which left 75 out of originally 100 fibril tracks in the nine spatial maps. The overlaid plots of the LOS velocity in Figure \ref{fig2} show that the locations determined from the temperature stratifications usually coincide with the locations of maximal flow speed even if we did not use them when defining the points in  temperature.

The azimuth angle of the flow was derived from the tangent to the fibril tracks at the spatial location of the photospheric point (see Figure \ref{fig1}). The azimuth of the fibril tracks and all other values of azimuth used were set to have 0\,deg to the right in each map while increasing in the counterclockwise direction.
\begin{figure}
\resizebox{8.8cm}{!}{\includegraphics{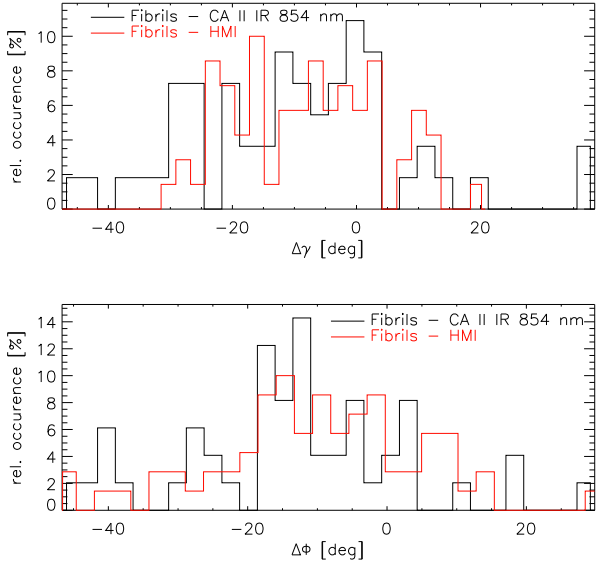}}
\caption{Histograms of the differences in inclination $\gamma$ (top panel) and azimuth $\phi$ (bottom panel) at the photospheric footpoints. Black (red) lines: difference between values the from the \ion{Ca}{ii} IR inversion (HMI) and the fibril tracks.}\label{histo2}
\end{figure}

We then extracted all relevant quantities from the various inversion results ($\gamma, \phi, B$) of the HMI, 854\,nm and 1565\,nm data at the location of the photospheric footpoints for each fibril in each spatial scan. The magnetic vector field and the flow direction were finally converted from the LOS reference frame (LOS RF) to the local reference frame (LRF) to provide the orientation relative to the solar surface.
% where $z$ indicates the local surface normal.
\section{Results} \label{secres}
\subsection{Magnetic field strength}
The top left panel of Figure \ref{histo1} shows the histograms of the magnetic field strength $B$ at the photospheric footpoints (FPs) in the HMI and 1565\,nm inversion. The majority of the FPs was located inside the penumbra (see Paper I). The field strength ranges from 0.5 to 1\,kG for HMI and from 1 to 1.5\,kG for the 1565\,nm inversion with average values of 0.8$\pm$0.2\,kG and 1.3$\pm$0.3\,kG, respectively. The HMI field strength is always only about 2/3 of that of the 1565\,nm inversion, as the scatter plot in the top right panel of Figure \ref{histo1} reveals. We confirmed this behavior by also looking at the scatter plot of $B$ for the complete umbra and penumbra of the sunspot. The ratio stayed about the same, while the average difference between the two values in the umbra was about 600\,G. In any case, the field strength at the inner FP is larger than at the outer end of the fibrils where no strong magnetic fields can be seen at all (see Figure \ref{fig1}).  
\subsection{Inclination}
The middle left panel of Figure \ref{histo1} shows the histogram of the inclination to the LOS derived from the HMI and \ion{Ca}{ii} IR inversion, and as derived from the temperature along the fibril tracks. The distributions roughly coincide, with the exception of a lack of small values below 30\,deg in HMI. The inclinations of the HMI and \ion{Ca}{ii} IR inversion match those derived from the fibril tracks within about $\pm 18$\,deg (right middle panel of Figure \ref{histo1} and top panel of Figure \ref{histo2}). The results from the inversion of \ion{Ca}{ii} IR scatter slightly more around the line of one-to-one correlation than those of HMI, but the linear polarization signals in the \ion{Ca}{ii} IR data are often small and affected by the noise level (Figure \ref{spec_exam}). The average inclination values and their standard deviations are listed Table \ref{table_results}. The average inclination to the LOS of about 50\,deg scales the flow velocities determined in Paper I up by about 50\,\%, which puts most values directly into the supersonic range. This comes in addition to the underestimation of the LOS velocities caused by having two different spectral components at the downflow points as discussed in Paper I.
\begin{figure}
\resizebox{8.cm}{!}{\includegraphics{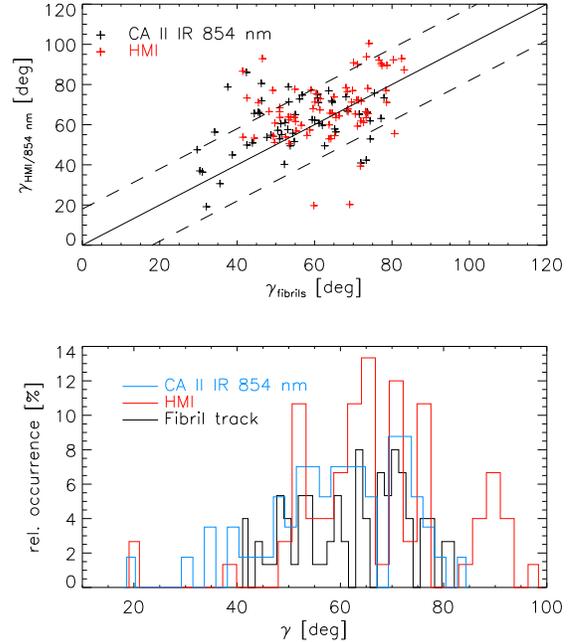}}
\caption{Inclination to the local vertical in the LRF. Top panel: scatter plot of the inclination $\gamma$ from HMI (red pluses) and the inversion of \ion{Ca}{ii} IR (black pluses) {\emph{vs.}} the inclination from the temperature along the fibril tracks. Bottom panel: histograms of the inclination from the temperature along the fibril tracks (black lines), HMI (red lines) and the  inversion of \ion{Ca}{ii} IR (blue lines).}\label{histo_lrf}$ $\\
\end{figure}

\subsection{Azimuth}
The corresponding plots for the azimuth in the bottom row of Figure \ref{histo1} demonstrate that the azimuth direction is the quantity that matches best across the different methods. The direction of the magnetic field azimuth in the HMI inversion corresponds to the direction of the flow fibrils as determined from the line-core intensity and line-core velocity maps within again a $\pm 18$\,deg range with an average value of mismatch below 10\,deg (bottom panel of Figure \ref{histo2} and Table \ref{table_results}). The azimuth in the \ion{Ca}{ii} IR inversion again scatters somewhat more, but only a few points lie outside the $\pm 18$\,deg range. 
\begin{table}
\caption{Average values and differences of $\gamma$ and $\phi$.}\label{table_results}
\begin{tabular}{ccccccccccc}
Average & fibril tracks & HMI & \ion{Ca}{ii} IR & \cr\hline\hline
$\gamma_{LOS RF}$ [deg]& $49\pm13$ & $57\pm15$ & $46\pm12$ \cr
$\gamma_{LRF}$ [deg]& $63\pm11$ & $68\pm15$ & $60\pm13$ \cr\hline
Difference & fibril tracks - HMI &\multicolumn{2}{c}{fibril tracks - \ion{Ca}{ii} IR}\cr\hline\hline
$\Delta\phi_{LOS RF}$  [deg]& $-7\pm19$ & \multicolumn{2}{c}{$-4\pm 32$} \cr
$\Delta\gamma_{LOS RF}$ [deg]& $-8\pm15$ & \multicolumn{2}{c}{$-9\pm 18$} \cr
$\Delta\gamma_{LRF}$ [deg]& $-5\pm15$ & \multicolumn{2}{c}{$-5\pm 15$} \cr
\end{tabular}
\end{table}

Figure \ref{histo2} shows the histograms of the differences between the three different approaches in inclination and azimuth in the LOS RF for completeness.

\subsection{Flow Angle in the Local Reference Frame}
We converted the flow angle and the orientation of the magnetic field from the LOS RF to the LRF following the approach in \citet[][Sect.~5.3.2]{beckthesis2006}. The top panel of Figure \ref{histo_lrf} shows a scatter plot of the inclination values in the LRF. The range of values from 30--90\,deg is slightly more compressed than in the LOS RF, with an average inclination to the local vertical of about 65\,deg (bottom panel of Figure \ref{histo_lrf} and Table \ref{table_results}). The estimate of 30--60\,deg for the angle to the local vertical in Paper I was derived visually from the plots of temperature along the fibril tracks such as in Figure \ref{fig2}, but without taking their 2:1 aspect ratio into account, which led to an underestimation of the inclination to the LOS, and hence to the LRF.  
%\section{Summary }\label{secsumm}
%We derived an estimate of the orientation and direction of flow channels harboring the inverse Evershed flow near their downflow points by three independent methods. We determined the azimuth angle and inclination of the flow channels from the direction of fibril tracks and the apparent inclination to the LOS of temperature structures. The other two approaches derived the direction of magnetic fields lines from an inversion of HMI and \ion{Ca}{ii} IR spectropolarimetric data. The inclination and azimuth values from all approaches match usually to about $\pm 15$\,deg. The match in the azimuth angle seems to be better than for the inclination (cf.~the lower two panels in the right column of Figure \ref{histo1}). The conversion to the LRF yielded an average inclination of the flow fibrils to the local surface normal of about 65\,deg at the downflow points. 
\section{Discussion}\label{secdisc}
We identified the tracks of flow fibrils from thermodynamic quantities, the line-core intensity and line-core velocity. The flow angle at the downflow points as derived from the direction of the fibril tracks and the apparent angle of the structures in the temperature inversion matches the magnetic field direction in the photosphere as derived from HMI data and from the polarization signal of the \ion{Ca}{ii} IR line within about $\pm 15$\,deg. That range lines up with the corresponding values for the deviation between fibrils and magnetic field vector found by \citet[][$\pm 16$\,deg]{asensio+etal2017} for \ion{Ca}{ii} IR and \citet[][$\pm 10$\,deg]{schad+etal2013} for \ion{He}{i} at 1083\,nm. This suggests that chromospheric flow fibrils derived from thermodynamic diagnostics trace magnetic field lines, at least near locations where they are returning down to the photosphere. 

The question whether chromospheric dark or bright fibrillar structures in general trace magnetic field lines goes back to a long-standing debate that was started based on H$\alpha$ observations \citep{foukal1971,foukal1971a,frazier1972,zirin1972,zirin1972a,foukal+zirin1972} and that was labeled the ``{\it Zirin-Frazier controversy}'' by \citet{cheng+etal1973}. The most remarkable statement about this controversy was made in \citet{cheng+etal1973}: ``{\it Any assumed relation between the H$\alpha$ fine structure and the chromospheric magnetic field can only be verified by actually measuring the magnetic field in the chromosphere. ... There is no short cut to obtaining the chromospheric field. We simply have to measure it.}'' %The version of \citet{cheng+etal1973} that is accessible online from the ADS is actually a scan of the printed version, and some unknown reader has already marked those lines in it.

More than four decades later, it might be worth to check whether solar physics has improved in that respect. The answer is unfortunately ``yes, but only partially''. The determination of the magnetic field vector from the polarization signal of the \ion{Ca}{ii} lines is strongly hampered by the small polarization amplitudes (see Figures \ref{fig1} and \ref{spec_exam}), especially for linear polarization, the formation of these lines with a strong contribution from the low to upper photosphere \citep{centeno2018} and the fact that structures such as flow fibrils are not necessarily opaque, but can be only a weak second component in a given spectrum (Paper I), apart from also being located at the upper end of the formation height range. Tracing the magnetic field lines outside of sunspots using \ion{Ca}{ii} polarimetry is thus nearly impossible in most cases. The approach of determining the azimuth and LOS inclination of apparent thermodynamic structures is useful to circumvent these limitations.
% of the polarimetry

Polarimetry of the chromospheric H$\alpha$ line has been used multiple times for the determination of the magnetic field in off-limb structures \citep[{\emph e.g.},][]{lopezariste+etal2005}, but on the disk it was rarely performed, maybe because of the words of caution about the interpretation in \citet{balasubramaniam+etal2004}. As a result, there currently is not much data of polarimetry in H$\alpha$ available at all just to investigate how well it would be suited for chromospheric magnetic field determination. 

The only line that has been widely established as being suited and that was used for the derivation of chromospheric magnetic fields is the \ion{He}{i} line at 1083\,nm \citep[\emph{e.g.},][]{ruedi+etal1995,lin+etal1998,lagg+etal2004,asensio+etal2008,bethge+etal2012,martinezgonzalez+etal2015}. 

Our current results on the locations of the downflow points of the inverse Evershed effect and the angle of the flow fibrils relative to the local vertical support the picture of the ``uncombed'' penumbra \citep[\emph{e.g.},][]{solanki+montavon1993,beck2011}. The flow angle of the IEF near the downflow points is not compatible with horizontal magnetic field lines, but indicates field lines that leave the photosphere and continue into the upper solar atmosphere. In the uncombed picture of the penumbra, nearly horizontal penumbral filaments harbor the regular Evershed flow. The inverse Evershed flow then would connect to other magnetic field lines that wrap around the horizontal filaments \citep{borrero+etal2008}.

The comparison of the magnetic field strength as derived from the inversion of HMI and 1565\,nm data showed a somewhat surprisingly large difference. Even if the near-IR lines at 1565\,nm form lower in the solar photosphere than the line used by HMI \citep{cabrera+bellot+iniesta2005}, the corresponding resulting field strength gradient of about 3\,G\,km$^{-1}$ in the umbra -- assuming a 600\,G difference in field strength and a 200\,km difference in formation height --  seems to be too large \citep[see][]{balthasar2018}. It might be worthwhile to acquire simultaneous spectropolarimetric observations of some near-IR line and the HMI line with high spectral resolution for a direct comparison \citep[see also][]{sainzdalda2017}. 

%Magnetograph data is by design intended to determine average magnetic flux and has not necessarily the spectral resolution to determine the magnetic field strength even if a formal inversion is run over it. The common usage and labeling of HMI inversion results as ``field strength'' might be overoptimistic.
%We plan to integrate our findings on the thermodynamic and magnetic properties of the IEF flow fibrils into a 3D volume model of such a fibril in the future. 
\section{Conclusions}\label{secconcl}
The flow fibrils that harbor the inverse Evershed flow (IEF) in the superpenumbra of sunspots are well aligned with the magnetic field orientation near their downflow points in the penumbra. The average field strength of about 1.3\,kG at those locations supports a siphon flow scenario as the driver of the IEF. The flow angle to the local surface normal of about 65\,deg indicates a connection of the IEF fibrils to more vertical penumbral magnetic field lines that wrap around horizontal regular Evershed flow filaments. This would imply that the EF and IEF are unrelated phenomena. Confirming a relation between intensity or velocity fibrils with the magnetic field direction derived from the analysis of spectropolarimetric observations in \ion{Ca}{ii} IR is nearly impossible outside of sunspots because of the intrinsic limitations on the \ion{Ca}{ii} IR polarization signal.
\begin{acknowledgements}
The Dunn Solar Telescope at Sacramento Peak/NM was operated by the National Solar Observatory (NSO). NSO is operated by the Association of Universities for Research in Astronomy (AURA), Inc.~under cooperative agreement with the National Science Foundation (NSF). HMI data are courtesy of NASA/SDO and the HMI science team. This work was supported through NSF grant AGS-1413686.
\end{acknowledgements}
\bibliographystyle{aa}
\bibliography{references_luis_mod}

\begin{thebibliography}{60}
\expandafter\ifx\csname natexlab\endcsname\relax\def\natexlab#1{#1}\fi

\bibitem[{{Asensio Ramos} {et~al.}(2017){Asensio Ramos}, {de la Cruz
  Rodr{\'{\i}}guez}, {Mart{\'{\i}}nez Gonz{\'a}lez}, \&
  {Socas-Navarro}}]{asensio+etal2017}
{Asensio Ramos}, A., {de la Cruz Rodr{\'{\i}}guez}, J., {Mart{\'{\i}}nez
  Gonz{\'a}lez}, M.~J., \& {Socas-Navarro}, H. 2017, \aap, 599, A133

\bibitem[{{Asensio Ramos} {et~al.}(2008){Asensio Ramos}, {Trujillo Bueno}, \&
  {Landi Degl'Innocenti}}]{asensio+etal2008}
{Asensio Ramos}, A., {Trujillo Bueno}, J., \& {Landi Degl'Innocenti}, E. 2008,
  \apj, 683, 542

\bibitem[{{Balasubramaniam} {et~al.}(2004){Balasubramaniam}, {Christopoulou},
  \& {Uitenbroek}}]{balasubramaniam+etal2004}
{Balasubramaniam}, K.~S., {Christopoulou}, E.~B., \& {Uitenbroek}, H. 2004,
  \apj, 606, 1233

\bibitem[{{Balthasar}(2018)}]{balthasar2018}
{Balthasar}, H. 2018, \solphys, 293, 120

\bibitem[{{Beck}(2006)}]{beckthesis2006}
{Beck}, C. 2006, PhD thesis, Albert-Ludwigs-University, Freiburg,
  \url{http://www.freidok.uni-freiburg.de/volltexte/2346/}

\bibitem[{{Beck}(2008)}]{beck2008}
{Beck}, C. 2008, \aap, 480, 825

\bibitem[{{Beck}(2011)}]{beck2011}
{Beck}, C. 2011, \aap, 525, A133

\bibitem[{{Beck} {et~al.}(2014){Beck}, {Choudhary}, \&
  {Rezaei}}]{beck+etal2014}
{Beck}, C., {Choudhary}, D.~P., \& {Rezaei}, R. 2014, \apj, 788, 183

\bibitem[{{Beck} {et~al.}(2015){Beck}, {Choudhary}, {Rezaei}, \&
  {Louis}}]{beck+etal2015}
{Beck}, C., {Choudhary}, D.~P., {Rezaei}, R., \& {Louis}, R.~E. 2015, \apj,
  798, 100

\bibitem[{{Beck} {et~al.}(2013){Beck}, {Rezaei}, \&
  {Puschmann}}]{beck+etal2013}
{Beck}, C., {Rezaei}, R., \& {Puschmann}, K.~G. 2013, \aap, 549, A24

\bibitem[{{Bellot Rubio} {et~al.}(2004){Bellot Rubio}, {Balthasar}, \&
  {Collados}}]{bellot+etal2004}
{Bellot Rubio}, L.~R., {Balthasar}, H., \& {Collados}, M. 2004, \aap, 427, 319

\bibitem[{{Bellot Rubio} {et~al.}(2003){Bellot Rubio}, {Balthasar}, {Collados},
  \& {Schlichenmaier}}]{bellot+etal2003}
{Bellot Rubio}, L.~R., {Balthasar}, H., {Collados}, M., \& {Schlichenmaier}, R.
  2003, \aap, 403, L47

\bibitem[{{Bethge} {et~al.}(2012){Bethge}, {Beck}, {Peter}, \&
  {Lagg}}]{bethge+etal2012}
{Bethge}, C., {Beck}, C., {Peter}, H., \& {Lagg}, A. 2012, \aap, 537, A130

\bibitem[{{Borrero} {et~al.}(2008){Borrero}, {Lites}, \&
  {Solanki}}]{borrero+etal2008}
{Borrero}, J.~M., {Lites}, B.~W., \& {Solanki}, S.~K. 2008, \aap, 481, L13

\bibitem[{{Borrero} {et~al.}(2011){Borrero}, {Tomczyk}, {Kubo},
  {Socas-Navarro}, {Schou}, {Couvidat}, \& {Bogart}}]{borrero+etal2011}
{Borrero}, J.~M., {Tomczyk}, S., {Kubo}, M., {et~al.} 2011, \solphys, 273, 267

\bibitem[{{Cabrera Solana} {et~al.}(2006){Cabrera Solana}, {Bellot Rubio},
  {Beck}, \& {del Toro Iniesta}}]{cabrerasolana+etal2006}
{Cabrera Solana}, D., {Bellot Rubio}, L.~R., {Beck}, C., \& {del Toro Iniesta},
  J.~C. 2006, \apjl, 649, L41

\bibitem[{{Cabrera Solana} {et~al.}(2005){Cabrera Solana}, {Bellot Rubio}, \&
  {del Toro Iniesta}}]{cabrera+bellot+iniesta2005}
{Cabrera Solana}, D., {Bellot Rubio}, L.~R., \& {del Toro Iniesta}, J.~C. 2005,
  \aap, 439, 687

\bibitem[{{Centeno}(2018)}]{centeno2018}
{Centeno}, R. 2018, \apj, 866, 89

\bibitem[{{Cheng} {et~al.}(1973){Cheng}, {Phillips}, \&
  {Wilson}}]{cheng+etal1973}
{Cheng}, C.-C., {Phillips}, K.~J.~H., \& {Wilson}, A.~M. 1973, \solphys, 29,
  383

\bibitem[{{Choudhary} \& {Beck}(2018)}]{choudhary+beck2018}
{Choudhary}, D.~P. \& {Beck}, C. 2018, \apj, 859, 139 (Paper I)

\bibitem[{{del Toro Iniesta} {et~al.}(2001){del Toro Iniesta}, {Bellot Rubio},
  \& {Collados}}]{deltoroiniesta+etal2001}
{del Toro Iniesta}, J.~C., {Bellot Rubio}, L.~R., \& {Collados}, M. 2001,
  \apjl, 549, L139

\bibitem[{{del Toro Iniesta} \& {Ruiz
  Cobo}(2016)}]{deltoroiniesta+ruizcobo2016}
{del Toro Iniesta}, J.~C. \& {Ruiz Cobo}, B. 2016, Living Reviews in Solar
  Physics, 13, 4

\bibitem[{{Esteban Pozuelo} {et~al.}(2016){Esteban Pozuelo}, {Bellot Rubio}, \&
  {de la Cruz Rodr{\'{\i}}guez}}]{pozuelo+etal2016}
{Esteban Pozuelo}, S., {Bellot Rubio}, L.~R., \& {de la Cruz Rodr{\'{\i}}guez},
  J. 2016, \apj, 832, 170

\bibitem[{{Evershed}(1909)}]{evershed1909}
{Evershed}, J. 1909, \mnras, 69, 454

\bibitem[{{Foukal}(1971{\natexlab{a}})}]{foukal1971}
{Foukal}, P. 1971{\natexlab{a}}, \solphys, 20, 298

\bibitem[{{Foukal}(1971{\natexlab{b}})}]{foukal1971a}
{Foukal}, P. 1971{\natexlab{b}}, \solphys, 19, 59

\bibitem[{{Foukal} \& {Zirin}(1972)}]{foukal+zirin1972}
{Foukal}, P. \& {Zirin}, H. 1972, \solphys, 26, 148

\bibitem[{{Franz} \& {Schlichenmaier}(2013)}]{franz+schliche2013}
{Franz}, M. \& {Schlichenmaier}, R. 2013, \aap, 550, A97

\bibitem[{{Frazier}(1972)}]{frazier1972}
{Frazier}, E.~N. 1972, \solphys, 24, 98

\bibitem[{{Gingerich} {et~al.}(1971){Gingerich}, {Noyes}, {Kalkofen}, \&
  {Cuny}}]{gingerich+etal1971}
{Gingerich}, O., {Noyes}, R.~W., {Kalkofen}, W., \& {Cuny}, Y. 1971, \solphys,
  18, 347

\bibitem[{{Harvey} \& {Harvey}(1973)}]{harvey+harvey1973}
{Harvey}, K. \& {Harvey}, J. 1973, \solphys, 28, 61

\bibitem[{{Jefferies} {et~al.}(1989){Jefferies}, {Lites}, \&
  {Skumanich}}]{jefferies+etal1989}
{Jefferies}, J., {Lites}, B.~W., \& {Skumanich}, A. 1989, \apj, 343, 920

\bibitem[{{Joshi} {et~al.}(2017){Joshi}, {Lagg}, {Hirzberger}, \&
  {Solanki}}]{joshi+etal2017}
{Joshi}, J., {Lagg}, A., {Hirzberger}, J., \& {Solanki}, S.~K. 2017, \aap, 604,
  A98

\bibitem[{{Khomenko} {et~al.}(2015){Khomenko}, {Collados}, {Shchukina}, \&
  {D{\'{\i}}az}}]{khomenko+etal2015}
{Khomenko}, E., {Collados}, M., {Shchukina}, N., \& {D{\'{\i}}az}, A. 2015,
  \aap, 584, A66

\bibitem[{{Kiess} {et~al.}(2014){Kiess}, {Rezaei}, \&
  {Schmidt}}]{kiess+etal2014}
{Kiess}, C., {Rezaei}, R., \& {Schmidt}, W. 2014, \aap, 565, A52

\bibitem[{{Kubo} {et~al.}(2008){Kubo}, {Lites}, {Shimizu}, \&
  {Ichimoto}}]{kubo+etal2008}
{Kubo}, M., {Lites}, B.~W., {Shimizu}, T., \& {Ichimoto}, K. 2008, \apj, 686,
  1447

\bibitem[{{Lagg} {et~al.}(2004){Lagg}, {Woch}, {Krupp}, \&
  {Solanki}}]{lagg+etal2004}
{Lagg}, A., {Woch}, J., {Krupp}, N., \& {Solanki}, S.~K. 2004, \aap, 414, 1109

\bibitem[{{Lin} {et~al.}(1998){Lin}, {Penn}, \& {Kuhn}}]{lin+etal1998}
{Lin}, H., {Penn}, M.~J., \& {Kuhn}, J.~R. 1998, \apj, 493, 978

\bibitem[{{L{\'o}pez Ariste} {et~al.}(2005){L{\'o}pez Ariste}, {Casini},
  {Paletou}, {Tomczyk}, {Lites}, {Semel}, {Landi Degl'Innocenti}, {Trujillo
  Bueno}, \& {Balasubramaniam}}]{lopezariste+etal2005}
{L{\'o}pez Ariste}, A., {Casini}, R., {Paletou}, F., {et~al.} 2005, \apjl, 621,
  L145

\bibitem[{{Mart{\'{\i}}nez Gonz{\'a}lez} {et~al.}(2015){Mart{\'{\i}}nez
  Gonz{\'a}lez}, {Manso Sainz}, {Asensio Ramos}, {Beck}, {de la Cruz
  Rodr{\'{\i}}guez}, \& {D{\'{\i}}az}}]{martinezgonzalez+etal2015}
{Mart{\'{\i}}nez Gonz{\'a}lez}, M.~J., {Manso Sainz}, R., {Asensio Ramos}, A.,
  {et~al.} 2015, \apj, 802, 3

\bibitem[{{Montesinos} \& {Thomas}(1997)}]{montesinos+thomas1997}
{Montesinos}, B. \& {Thomas}, J.~H. 1997, \nat, 390, 485

\bibitem[{{Pesnell} {et~al.}(2012){Pesnell}, {Thompson}, \&
  {Chamberlin}}]{pesnell+etal2012}
{Pesnell}, W.~D., {Thompson}, B.~J., \& {Chamberlin}, P.~C. 2012, \solphys,
  275, 3

\bibitem[{{Rempel}(2012)}]{rempel2012}
{Rempel}, M. 2012, \apj, 750, 62

\bibitem[{{Rempel}(2015)}]{rempel2015}
{Rempel}, M. 2015, \apj, 814, 125

\bibitem[{{Rezaei} {et~al.}(2006){Rezaei}, {Schlichenmaier}, {Beck}, \& {Bellot
  Rubio}}]{reza+etal2006}
{Rezaei}, R., {Schlichenmaier}, R., {Beck}, C., \& {Bellot Rubio}, L.~R. 2006,
  \aap, 454, 975

\bibitem[{{Rueedi} {et~al.}(1995){Rueedi}, {Solanki}, \&
  {Livingston}}]{ruedi+etal1995}
{Rueedi}, I., {Solanki}, S.~K., \& {Livingston}, W.~C. 1995, \aap, 293, 252

\bibitem[{{Ruiz Cobo} \& {Asensio Ramos}(2013)}]{ruizcobo+asensio2013}
{Ruiz Cobo}, B. \& {Asensio Ramos}, A. 2013, \aap, 549, L4

\bibitem[{{Ruiz Cobo} \& {del Toro Iniesta}(1992)}]{cobo+toroiniesta1992}
{Ruiz Cobo}, B. \& {del Toro Iniesta}, J.~C. 1992, \apj, 398, 375

\bibitem[{{Sainz Dalda}(2017)}]{sainzdalda2017}
{Sainz Dalda}, A. 2017, \apj, 851, 111

\bibitem[{{Schad} {et~al.}(2013){Schad}, {Penn}, \& {Lin}}]{schad+etal2013}
{Schad}, T.~A., {Penn}, M.~J., \& {Lin}, H. 2013, \apj, 768, 111

\bibitem[{{Scherrer} {et~al.}(2012){Scherrer}, {Schou}, {Bush}, {Kosovichev},
  {Bogart}, {Hoeksema}, {Liu}, {Duvall}, {Zhao}, {Title}, {Schrijver},
  {Tarbell}, \& {Tomczyk}}]{scherrer+etal2012}
{Scherrer}, P.~H., {Schou}, J., {Bush}, R.~I., {et~al.} 2012, \solphys, 275,
  207

\bibitem[{{Schlichenmaier} \& {Schmidt}(2000)}]{schliche+schmidt2000}
{Schlichenmaier}, R. \& {Schmidt}, W. 2000, \aap, 358, 1122

\bibitem[{{Siu-Tapia} {et~al.}(2017){Siu-Tapia}, {Lagg}, {Solanki}, {van
  Noort}, \& {Jur{\v c}{\'a}k}}]{siutapia+etal2017}
{Siu-Tapia}, A., {Lagg}, A., {Solanki}, S.~K., {van Noort}, M., \& {Jur{\v
  c}{\'a}k}, J. 2017, \aap, 607, A36

\bibitem[{{Socas-Navarro} {et~al.}(2006){Socas-Navarro}, {Elmore}, {Pietarila},
  {Darnell}, {Lites}, {Tomczyk}, \& {Hegwer}}]{socasnavarro+etal2006}
{Socas-Navarro}, H., {Elmore}, D., {Pietarila}, A., {et~al.} 2006, \solphys,
  235, 55

\bibitem[{{Solanki}(2003)}]{solanki2003}
{Solanki}, S.~K. 2003, \aapr, 11, 153

\bibitem[{{Solanki} \& {Montavon}(1993)}]{solanki+montavon1993}
{Solanki}, S.~K. \& {Montavon}, C.~A.~P. 1993, \aap, 275, 283

\bibitem[{{Thomas}(1988)}]{thomas1988}
{Thomas}, J.~H. 1988, \apj, 333, 407

\bibitem[{{Vrabec}(1974)}]{vrabec1974}
{Vrabec}, D. 1974, in IAU Symposium, Vol.~56, Chromospheric Fine Structure, ed.
  R.~G. {Athay}, 201

\bibitem[{{Zirin}(1972{\natexlab{a}})}]{zirin1972}
{Zirin}, H. 1972{\natexlab{a}}, \solphys, 22, 34

\bibitem[{{Zirin}(1972{\natexlab{b}})}]{zirin1972a}
{Zirin}, H. 1972{\natexlab{b}}, \solphys, 26, 145

\end{thebibliography}

\end{document}